\documentstyle[psfig, art11, conf_iap]{article}

\begin{document}

\psnoisy

\heading{Modelling the spectro-photometric and chemical evolution of Low 
Surface Brightness spiral galaxies}

\author{L.B. ~van den Hoek$^{1}$ and W.J.G. ~de Blok$^{2}$}
        {$^{1}$ Astronomical Institute 'Anton Pannekoek',
        University of Amsterdam, Kruislaan 403, NL 1098 SJ Amsterdam, 
        The Netherlands \\ $^{2}$Kapteyn Astronomical Institute, 
	P.O. Box 800, NL 9700 AV Groningen, The Netherlands}

\bigskip

\begin{abstract}{\baselineskip 0.4cm 
We investigate the star formation history and chemical evolution of Low 
Surface Brightness (LSB) spiral galaxies by means of their observed
spectro-photometric and chemical properties. We present preliminary 
results for Johnson-Cousins UBVRI magnitudes and stellar [O/H] abundance
ratios using a galactic chemical evolution model incorporating a detailed 
metallicity dependent set of stellar input data covering all relevant stages 
of stellar evolution. Comparison of our model results with observations 
confirms the idea that LSB galaxies are relatively unevolved systems. 
However, we argue that recent and ongoing massive star formation plays an
important role in determining the colours of many LSB spirals. 
We briefly discuss these results in the context of the spectral evolution of 
spiral galaxies in general.}
\end{abstract}

\section{Observations}

Low Surface Brightness spiral galaxies (LSBs) are generally late-type 
galaxies with blue central surface brightnesses fainter than $\sim$23 mag 
arcsec$^{-2}$. They usually display ill-defined spiral arms in the optical 
and contain only a few bright H{\sc ii} regions.
Comparison of the chemical and optical properties of LSBs and 
HSB spirals reveals that LSBs are, on average, less luminous, bluer, 
metal-deficient, and more extended than HSBs \cite{BHB}, \cite{MSB}.
Furthermore, LSBs have lower H{\sc i} surface densities and 
smaller dynamical masses within their optical radii than their HSB 
counterparts \cite{BMH}.
In this paper, we investigate the spectro-photometric and chemical 
properties of LSBs by means of a detailed galactic evolution model.

\section{Model assumptions}

We concentrate on the stellar contribution to 
the total galaxy luminosity in a given passband (other contributions are 
neglected). For a given star formation history (SFR), we compute the chemical 
enrichment of a model galaxy by successive generations of evolving stars. 
To derive the stellar luminosity in a given passband at a given age, we 
use an up-to-date metallicity dependent set of theoretical stellar isochrones 
as well as a library of spectro-photometric data. The 
spectro-photometric properties of the model galaxy are calculated by 
integrating the stellar luminosities at a given galactic age 
weighted with the SFR at the time these stars were born \cite{HB}.

\section{Results}

We show in Fig. 1 model results for an exponentially decaying 
SFR$\propto \exp (-t/ \tau_{\rm sfr})$ with $\tau_{\rm sfr}$ = 5 Gyr 
(appropriate for the Galactic disk).
We followed the chemical and spectro-photometric evolution of the model galaxy 
during the last 14 Gyr. A power-law IMF (slope $-$2.35) 
and initial stellar mass limits of 0.1 and 60 \ms have been assumed. \\

\leftline{\psfig{figure=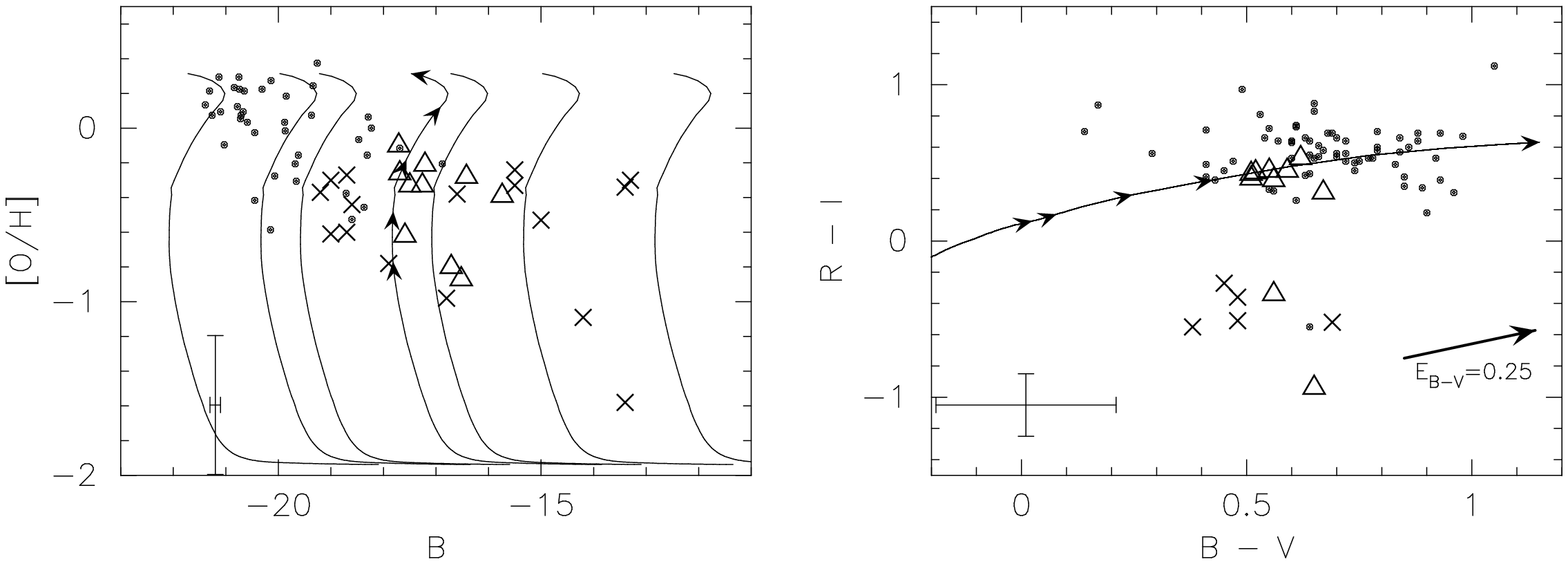,height=7.cm,width=16.cm,angle=270.}}
\vspace{-0.1cm} \begin{figure*}[h]
\caption[]{Resulting chemical and spectro-photometric evolution in the 
case of an exponentially decaying SFR ({\em solid curves}).
{\em Left panel}: predicted [O/H] vs. {\em B} relation.
Curves correspond to different initial galaxy masses in the range 10$^{7}$ 
to 5 10$^{10}$ \mss. For the 10$^{9}$ \ms model, arrows indicate evolution 
times of 1, 2, 4, 8, and 14 Gyr, respectively. {\em Right panel}: 
predicted {\em R}$-${\em I} vs. {\em B}$-${\em V} relation (independent of 
initial galaxy mass).
Corrections for internal extinction will shift the model predictions in the 
direction indicated by the arrow.
Symbols refer to the following galaxy types: face-on HSB spirals
(dots \cite{ROE}), LSB spiral galaxies (triangles \cite{BHB}), and dwarf 
irregulars (crosses \cite{ISR}). Observational errors are indicated in the 
bottom left corner.}
\end{figure*}

\section{Discussion and conclusion}

We find that the colours and abundances of HSBs can be reasonably well 
explained by the decaying SFR model ending at a current gas-to-total
mass-ratio of $\mu =$0.1 (deviation from individual HSBs is partly
due to the effects of dust-extinction).
However, colours and [O/H] abundances predicted 
by the same models are inconsistent with observational data of LSBs, 
within the B-magnitude range appropriate for LSBs.
These models clearly demonstrate that LSBs must have experienced their 
dominant episode of star formation much later in their evolution than HSBs.
Furthermore, LSBs show [O/H] abundances considerably smaller than 
expected for values of $\mu \lsim$0.15 which are suggested by the observations. 
This may indicate that: 1) LSBs experience predominantly star formation out 
of metal-deficient material not participating in the chemical enrichment of 
the disk, and/or 2) LSBs have much larger ratios $\mu$ than observed.
Since detailed observations appear to rule out the second possibility 
\cite{BMH}, we conclude that metal-deficient gas infall is important in LSBs. 
In this case, recent and ongoing star formation in LSBs is required to 
maintain $\mu$ as low as 0.15.

\begin{footnotesize}

\end{footnotesize}
\vfill
\end{document}